# Does the Earth show up an impending mega-earthquake?

Leontina L. Romashkova[1] and Vladimir G. Kossobokov[1,2].

[1] International Institute of Earthquake Prediction Theory and Mathematical Geophysics, Russian Academy of Sciences, Moscow, Russia
[2] Institute de Physique du Globe de Paris, France

**Abstract**

In line of the intermediate-term monitoring of seismic activity aimed at prediction of the world largest earthquakes the seismic dynamics of the Earth's lithosphere is analysed as a single whole, which is the ultimate scale of the complex hierarchical non-linear system. The present study demonstrates that the lithosphere does behave, at least in intermediate-term scale, as non-linear dynamic system that reveals classical symptoms of instability at the approach of catastrophe, i.e., mega-earthquake. These are: (i) transformation of magnitude distribution, (ii) spatial redistribution of seismic activity, (iii) rise and acceleration of activity, (iv) change of dependencies across magnitudes of different types, and other patterns of collective behaviour. The observed global scale seismic behaviour implies the state of criticality of the Earth lithosphere in the last decade.

## Introduction

Correlation in earthquake occurrence at long distances is a prominent feature of seismic dynamics investigated in many seismological studies. Such correlation appears as simultaneous changes of seismic activity within a large number of neighbouring regions (Keilis-Borok & Malinovskaya, 1964; Dobrovolsky et al., 1979; Sykes & Jaume 1990; Zaliapin et al. 2002), migration of the earthquakes along the fault zones (Mogi 1968, 1985; Vil'kovich & Shnirman 1983; Wallace 1987), long-range interaction of the earthquakes (Prozorov & Schreider 1990; Kagan & Jackson 1991, Shebalin et al., 2000), global spatial-and-temporal patterns in the seismic energy release (Benioff 1951; Mogi 1979; Romanowicz 1993; Bufe & Perkins 2005) and other. Although definite physical mechanism of stress and strain redistribution over such long distances is still unknown, the long-range correlation of earthquakes might be attributed to large- or even global-scale processes taking place in the lithosphere (Barenblatt et al., 1983; Press & Allen, 1995). Another, more general approach regards long-range correlation as a common feature of complex non-linear hierarchical dynamic systems (Keilis-Borok, 1990, 2002; Sornette & Sammis, 1995; Turcotte et al., 2000), which example the lithosphere of the Earth represents.

There is growing evidence that the process of earthquake generation is not localised about its future source but involves an area that exceeds the size of forthcoming earthquake by ten or more times (Dobrovolsky et al., 1979; Sadovsky, ed. 1986; Keilis-Borok 1990, 2002; Press & Allen, 1995; Bowman et al. 1998). Accordingly, manifestations of an earthquake approach may also come from territory that wide. This concept provided foundation for a series of earthquake prediction algorithms (see Keilis-Borok & Soloviev, 2003, for the list and descriptions). The authors regard a lithosphere as a complex non-linear hierarchical dynamic system where strong earthquakes are critical phenomena, different for different scales of the hierarchy. By using a pattern recognition technique they search for symptoms of the system instability within rather wide territory around prospective earthquake source.

The linear dimension $L$ of the investigation area depends on the magnitude $M_0$ of the target earthquake; usually, $L(M_0)$ is about 5-10 times more than its rupture length. For example, in the M8 algorithm (Keilis-Borok & Kossobokov, 1990), $L(8.0)$ is about 1300 km in diameter. Naturally, the stronger is the target earthquake the wider is the territory involved in the process of its preparation. The algorithms being tested in real-time experiments demonstrate high predictive ability both at the global and regional scales (Kossobokov et al. 1999; Rotwain & Novikova, 1999; Romashkova & Kossobokov, 2005) that, in fact, is solid indirect evidence in support of the extended earthquake preparation area.

The recent devastating mega-thrust December 26, 2004, magnitude 9.0 earthquake ripped the subduction zone over 1500 km from off west coast of Northern Sumatra to Andaman Islands. The event is among the top ten instrumentally recorded earthquakes of the last hundred years. Following the concept described above one can deduce that preparation of the 2006 Sumatra-Andaman earthquake might have involved the territory of about 7500–15000 km in linear dimension. This size exceeds the radius of the Earth and is comparable with the extent of the Earth's hemisphere. It suggests that the generation of such a mega-earthquake is likely to have a world-wide scale, rather than to involve interaction of a confined regional fault system. Therefore it is not unreasonable to search for symptoms of instability at the approach of the mega-earthquake, and hence some premonitory phenomena, by analysing the lithosphere as a single whole, which represents the ultimate scale of the complex Earth's hierarchy. The premonitory phenomena may express themselves in various changes of basic integral characteristics of the global seismic activity or other geophysical parameters, as well as to appear at different time scales that are unknown *a priori*.

The basic types of the seismic premonitory phenomena are listed in Keilis-Borok, (1990, 2002). They reflect changes in the rate and regularity of earthquake occurrence, their territory distribution, clustering in space and time, and relation between earthquakes from different magnitude ranges. The premonitory phenomena is captured by different seismicity patterns within a given territory, time interval and magnitude range. The time scale of the patterns varies from $10^2$ years to hours that implies rough subdivision of the prediction problem into *large-*, *intermediate-* and *short-term*. The magnitude range can be different for different patterns, so sometimes several ranges are analysed in parallel to ensure reliable results (Kossobokov et al., 1999). Due to the complex nature of seismicity a use of combination of the patterns, even if correlated by definition, often yields better prediction than monitoring of the individual one (Keilis-Borok & Soloviev, 2003).

In present work, by analysing the world earthquake catalogues, we intend to find the *global-scale* (involving the whole lithosphere) *intermediate-term* (formed within several years) *seismic* precursory phenomena of mega-earthquake. To investigate the problem we analyse the temporal variations of the frequency-magnitude distribution (hereinafter FM) of the global seismicity for several layers of the Earth depth. Such approach allows us to control emergence of precursory patterns related to variation of earthquake rates in different magnitude ranges, as well as to redistribution of the seismic activity in depth.

The two world-wide catalogues, i.e., the National Earthquake Information Center Global Hypocenters Database (GHDB, 1989) and the Harvard Centroid Moment Tensor catalogue (Ekstrom et al., 2005, and references therein), are used. It is of common knowledge that it is of great importance, for temporal analysis in particular, to have a clear understanding on homogeneity and completeness of the data source in the space-time-magnitude domain under study. Therefore, to determine the data for further investigation, we give special attention to the detailed analysis of the catalogues. The necessary precondition implies using the uniform and homogeneous magnitude scale throughout extended range (of at least several magnitude units) and the stable completeness of the data covering the territory of interest over a long period of time. The preliminary data analysis is described in the first paragraph. Then we apply temporal frequency-magnitude-depth analysis to the uniform data and draw inferences from empirical observations. Finally, we discuss the results of analysis in connection with those of the previous studies and make conclusions.

## Preliminary data analysis

The first considered is the US Geological Survey, **National Earthquake Information Center Global Hypocenters Database** (GHDB, 1989) with duplicates identified and removed by automatic procedure of P. Shebalin (1992), routinely updated through 2004 from NEIC Preliminary Determination of Epicenters data (PDE) available on-line (http://neic.usgs.gov/neis/data_services/ftp_files.html). It is referred hereinafter as NEIC. Every event in NEIC is supplied with the original time, latitude, longitude, depth and up to four magnitude determinations $M1$, $M2$, $M3$ and $M4$, along with other information. First and second magnitudes are occupied by $mb$ and $M_S$ correspondingly. Since mid-1960ies when the World Wide Standard Seismograph Network was installed $M1$ and $M2$ are the average determinations of $mb$ and $M_S$ reported by seismic stations. The other two magnitudes $M3$ and $M4$ report estimates of different type attributed to different authority or agency. Since the beginning of 1993 $M3$ and $M4$ host for the most part the moment magnitudes determined at USGS or at Harvard University.

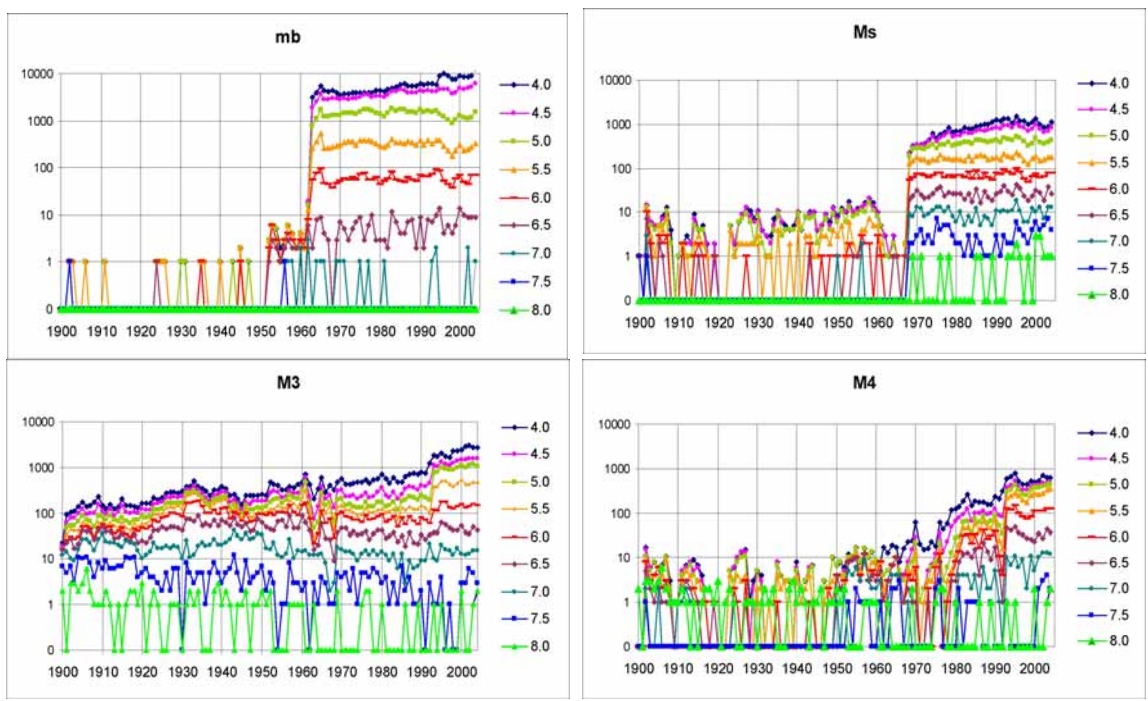

**Figure 1**. Annual number of earthquakes in NEIC catalogue by time and magnitude, 1900-2004. Different coloured lines stacked from higher ranges provide the number of earthquakes above a certain magnitude threshold $M_{threshold}$, with the step of 0.5 magnitude. The diagrams correspond to the four magnitudes in the NEIC: a) $mb$ ($M1$), b) $M_S$ ($M2$), c) $M3$ and d) $M4$.

Fig. 1 displays the temporal evolution of the annual number of earthquakes of different magnitude ranges for each of the four magnitudes *M1, M2, M3, M4* reported in NEIC. Different ranges of 0.5-magnitude width are given in different colour. The ordinate is in logarithmic scale to facilitate recognising whether the frequency-magnitude distribution fits the Gutenberg-Richter relation (Gutenberg & Richter, 1954; hereinafter G-R): in fact, a near-uniform distance between adjacent FM plots corresponds to a power-law distribution of earthquake sizes. Fig. 1 illustrates the great difference of completeness and stability of the reported magnitudes. Specifically, the magnitude *mb* (*M1*) is quite uniformly presented starting from 1963. Similarly, the magnitude $M_S$ (*M2*) appears uniform from 1968, although the number of *mb* determinations in NEIC is about ten times larger than that of $M_S$. The magnitudes *mb* and $M_S$ below 4.5 are essentially incomplete in all times to the present. Moreover, the number of magnitude 4.5 events grows gradually and reaches the level of completeness at about 1995. Thus, when analysing seismic activity in 1969-2004 making use of *mb* and $M_S$ the level of completeness should be set at the common value about 5.0 or larger. Note that the body-wave magnitude scale *mb* is known for a saturation effect, i.e., for being almost insensitive to distinguish and order large size earthquakes, which is evident from the increasingly larger widths for magnitudes *mb* above 6.0 and the absence of *mb*=7.5 or more in Fig. 1. The *M3* and *M4* although presented since the beginning of the last century are not uniform in time. In fact, several evident changes in the level of completeness can be observed in Fig. 1 (e.g., 1931, mid-1960s, 1994) that is explained by the diversity of the magnitude scales appearing in these two positions of NEIC. Therefore, the use of *M3* and *M4* without a multitude of proper calibrations in an analysis of seismic variability seems futile at least before 1993.

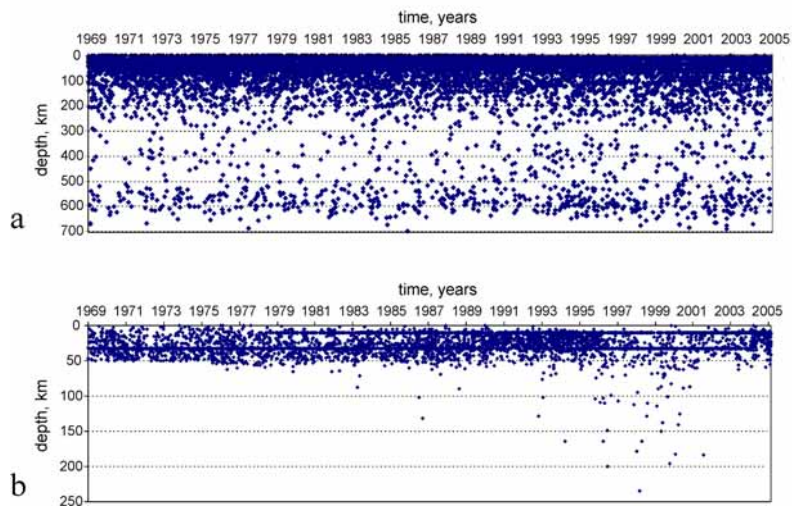

**Figure 2.** Depth versus time distribution of earthquakes from NEIC: a) body-wave magnitude *mb* ≥ 5.5; b) surface-wave magnitude $M_S$ ≥ 5.5, 1969-2004.

Most of the events in NEIC (about 99%) both shallow and deep holds *mb* determination (Fig. 2a). Fig. 3 shows frequency-depth distributions plotted for earthquakes in 1969-2004 from different ranges of *mb*. The ordinate is in logarithmic scale, so that a vertical slice of the diagram represents the FM distribution of events at a particular depth. One can see that the frequency-depth distributions are very similar in all magnitude ranges: The majority of earthquakes occur within the upper layer of the lithosphere; going deeper a near exponential decrease of the rate is observed down to 300 km; in between 300 and 500 km seismic activity has roughly the same rate; further down, at depths of 500–600 km there is a certain increase of the seismic rate followed by sharp falloff to zero at about 700 km.

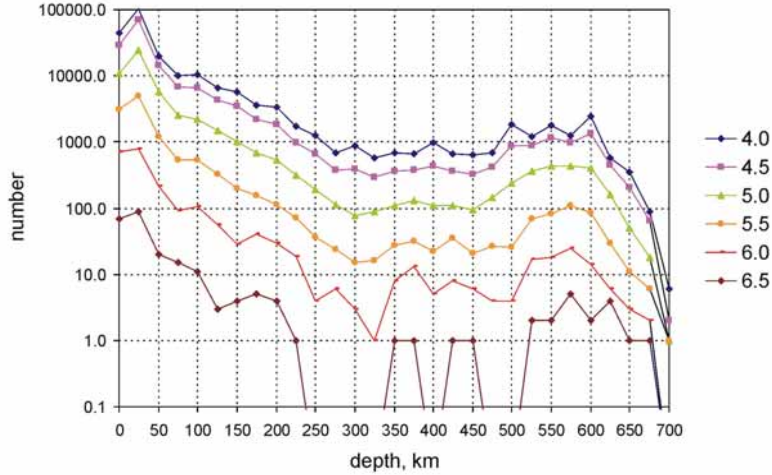

**Figure 3**. Frequency-depth distributions of the earthquakes with *mb* magnitude for different magnitude ranges, 1969-2004. Different coloured lines provide the number of earthquakes above a certain magnitude threshold $M_{threshold}$, with the step of 0.5 magnitude. The width of the depth sorting bins is 25 km.

It is of common knowledge that, although some exceptions exist, magnitude $M_S$ is determined mostly for shallow earthquakes. Indeed, Fig. 2b, shows explicitly that the depth, $h$, of an overwhelming majority of earthquakes with reported $M_S$ in NEIC does not exceed 50 km. More deep earthquakes are uncommon and their appearance in time is irregular. Furthermore often, when automatic determination of the depth is problematic, earthquakes are used to be put by seismologists at a few standard depths, e.g., for shallow ones at 10 km (19% of all events) or 33 km (34%).

Thus taking into account the above-mentioned limitations we consider for the analysis here the period 1969-2004 and the two NEIC magnitudes *mb* (*M1*) and $M_S$ (*M2*) (Table 2). The analysis based on *mb* will involve earthquakes from all depths along with their subdivisions into upper ($h \leq 100$ km), intermediate (100 km $< h \leq$ 300 km), deep (300 km $< h \leq$ 500 km) and ultra-deep (h $>$ 500 km) ones. The analysis based on $M_S$ will be performed for all shallow earthquakes with $h \leq 50$ km.

Next we analyse for different magnitude ranges the annual percentage *P* of the records holding *mb* and $M_S$ in NEIC. The results are summarised in Table 1 and Fig. 4. It is evident that the *mb* magnitude is the dominant source of the earthquake size in NEIC: it is reported for 99% of all registered earthquakes in all magnitude ranges down to magnitude 5.0, and the high percentage remains persistent in 1969-2004. On the contrary, the $M_S$ is reasonably complete (*P* $>$ 95% when averaged over the entire time of analysis) for magnitude 6.0 and above only. For lower range of magnitude 5.5–6.0 *P* reaches about constant level of 90% after 1980. For 5.0–5.5 the percentage of the $M_S$ is very unstable, it varies between 20% and 80%, gradually increases with time with average about 50%. Therefore, we conclude that seismic activity in 1969-2004 could be described with a great confidence either by earthquakes of the *mb* $\geq$ 5.0 or those of the $M_S \geq$ 6.0 (Table 1). At the same time we keep in mind the saturation of the *mb* magnitude scale and the restricted use of the $M_S$ magnitude scale for sizing, almost exclusively, shallow earthquakes.

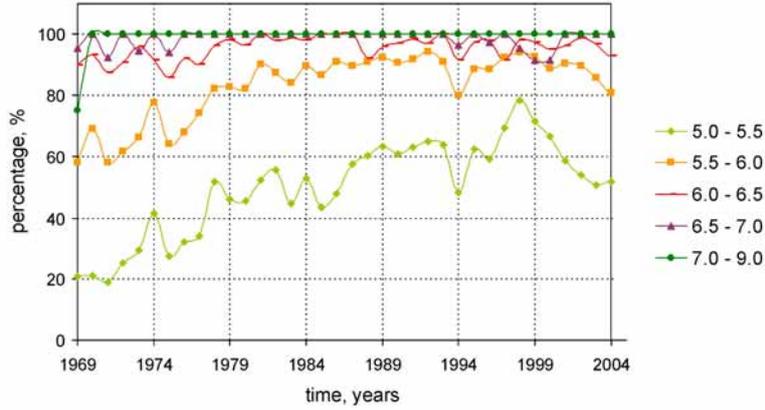

**Figure 4**. Annual percentage of the NEIC records holding $M_S$ magnitude for different magnitude ranges, 1969-2004. Earthquakes with depth≤50 km are considered. Different coloured lines correspond to the number of earthquakes within certain $M_{max}$ ranges.

**Table 1**. Representativeness of *mb* and *Ms* magnitudes in the NEIC, 1969-2004. *<P>* - average annual percentage of the records with a particular magnitude, in %; *Sigma* – standard deviation of *P*, in %. All calculated for different magnitude ranges defined by $M_{low} \leq M_{max} \leq M_{high}$, where $M_{max}$ is maximum of all magnitudes presented for the event. For *mb* magnitude the earthquakes of all depth are considered, for *Ms* – the shallow events only.

| $M_{max}$ range | *mb* | | *Ms* | |
|---|---|---|---|---|
| | *<P>* | σ | *<P>* | σ |
| 7.0 – 9.0 | 98.7 | 3.9 | 99.3 | 4.1 |
| 6.5 – 7.0 | 99.6 | 1.2 | 98.6 | 2.7 |
| 6.0 – 6.5 | 99.5 | 1.3 | 95.5 | 3.8 |
| 5.5 – 6.0 | 99.6 | 1.0 | 83.0 | 10.7 |
| 5.0 – 5.5 | 99.6 | 0.5 | 49.8 | 15.2 |

The Harvard University group compiles on a regular basis centroid-moment tensor solutions for the entire Globe (Ekstrom et al., 2005). For each seismic event their **Centroid Moment Tensor catalogue**, CMT catalogue provides information including the origin time, latitude, longitude, depth, magnitudes *mb* and $M_S$, and scalar moment $M_0$. Following Kanamori (1977) the scalar moment is usually recalculated into the so-called moment magnitude *Mw*:

$$Mw = 2/3 \cdot \log_{10} M_0 - 10.73 \qquad (1)$$

The *mb* and $M_S$ magnitudes in CMT are given as reported by other agencies (usually, NEIC or ISC) to the date of centroid-moment tensor solution. Therefore, near all of them present preliminary estimates released by NEIC as Quick Earthquake Determinations subject to revision in the final NEIC GHDB. Under the circumstances we do not consider these magnitudes and use *Mw* as the CMT primary measure of seismic event size.

Fig. 5 displays the temporal evolution of the annual number of earthquakes of different magnitude ranges of *Mw*. Apparently, starting from 1976 the data on earthquakes with $Mw \geq 6.0$ are reasonably complete and homogeneous. From 1977 the magnitude level of completeness rises to 5.5 and approximates to 5.0 after mid-1990ies. The analysis of the frequency-depth and depth-time distributions reveals patterns similar to those observed in NEIC for *mb*. Thus the analysis based on CMT *Mw* will involve earthquakes in 1977-2004 from all depths and in the same depth ranges as defined for NEIC magnitude *mb* (Table 2).

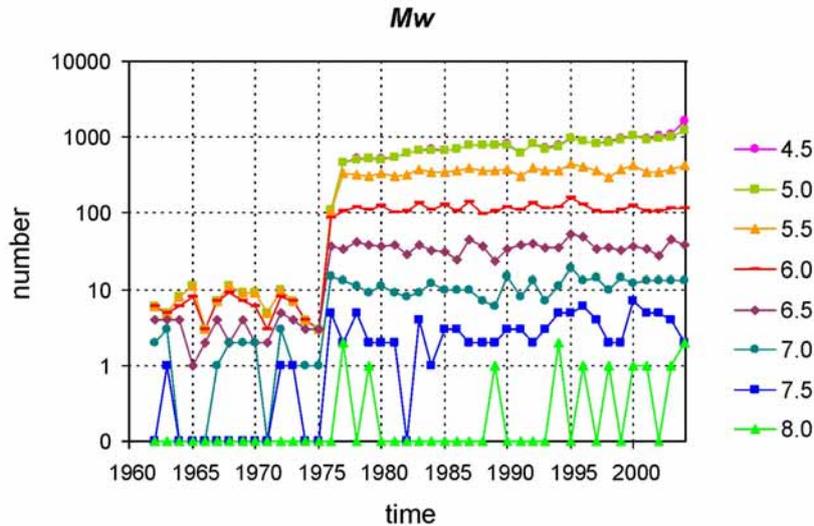

**Figure 5.** Annual number of earthquakes in the CMT catalogue by time and magnitude $M_W$, 1960-2004. Different coloured lines stacked from higher ranges provide the number of earthquakes above a certain magnitude threshold with the step of 0.5 magnitude.

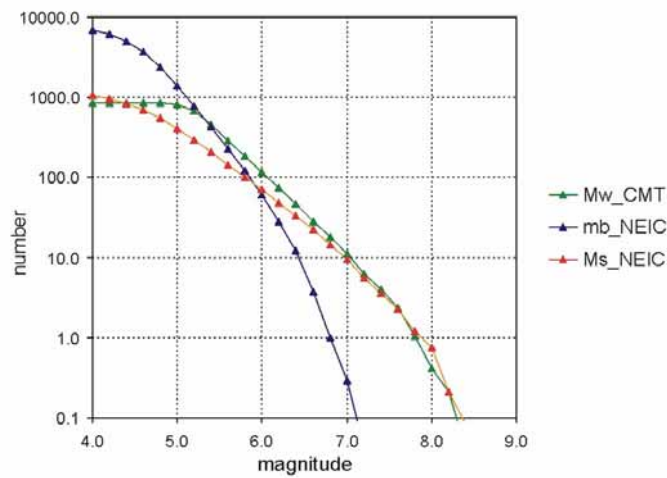

**Figure 6.** Annually averaged cumulative frequency-magnitude distributions for *mb* (blue), $M_S$ (red) and $M_W$ (green), 1981-2004

Fig. 6 represents the empirical cumulative annual average FM distributions of the *mb*, $M_S$, and *Mw* in 1981-2004. (Specifically, the three FM distributions integrate the information presented in Figs 1a, b and 5.) None of the three distributions fit a straight line on the entire magnitude range. Apparently, the curves bend on the left due to incompleteness of data records, while on the right due to magnitude saturation. The graph for *mb* is much steeper than either for $M_S$ or *Mw*. Evidently, the completeness of the NEIC data on *mb* and $M_S$ is better than that of the CMT *Mw*. For obvious reasons the number of *mb* determinations is about ten times larger than those of $M_S$ or *Mw*.

Table 2 summarises information of the data being used in present work for temporal analysis.

**Table 2**. The data have been used in the current work for temporal analysis.

| Data | Time period | Depth intervals, km |
|---|---|---|
| NEIC, *mb* | 1969 - 2004 | 0 – 100, 100 - 300, 300 - 500, 500 -700 |
| NEIC, *Ms* | 1969 - 2004 | 0 - 50 |
| CMT, *Mw* | 1977 - 2004 | 0 – 100, 100 - 300, 300 - 500, 500 -700 |

Finally let us explore in brief the question of using the magnitude of mixed type in temporal analysis of seismic activity. For example, $M_{max}$ defined as the maximum of all (or selected) values of magnitudes presented in NEIC may be useful in some applications. Such a composite magnitude allows bypassing the incompleteness of the $M_S$–type scale at low magnitudes and the saturation of the *mb*–type scale at high magnitudes. On the other hand the temporal behaviour of $M_{max}$ critically depends on the time homogeneity and stability of primary magnitudes. In case of piecewise or unstable data (that is usual for seismic catalogues) $M_{max}$ switches from one magnitude type to another depending on their presence in the catalogue. The difference in magnitude scales (see Fig. 6) may result in hectic temporal variations not related to the actual seismic activity (for an example, see Fig. 7). It is rather evident that before 1993 the contributions of the four magnitudes *mb*, $M_S$, *M3,* and *M4* into the cumulative number of events with $5.5 \leq M_{max} \leq 6.0$ were proportional. The proportion did break into a new one after the moment magnitudes were introduced and eventually became dominant contributors to the values of $M_{max}$. Similar graphs are obtained for higher 0.5-magnitude bins. The analysis of the man-made changes in NEIC and other seismic data is interesting but lies outside the scope of the present paper. Hereinafter we rely on persistent procedures of magnitude determinations at seismic stations contributing to average values of *mb* and $M_S$.

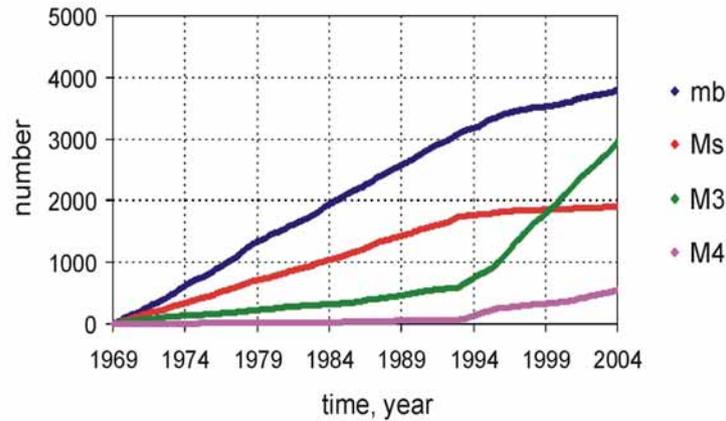

**Figure 7.** The contributions of the four NEIC magnitudes into the cumulative number of events with $5.5 \leq M_{max} \leq 6.0$. Different coloured lines represent events with $M_{max}$ equals *mb*, $M_S$, *M3* or *M4* correspondingly.

# Temporal analysis of the global frequency-magnitude-depth distribution

Thus, to investigate the temporal behaviour of the global seismic activity we analyse the variations of frequency-magnitude distribution computed in the four consecutive layers of depth. The analysis is based on the data from global catalogues summarised in Table 2. All earthquakes are considered without fore-, main- and after-shock discriminations. The geographical coordinates are not taken into account. The origin time, magnitude and depth of the earthquakes is the only information in use.

**NEIC, *mb*.** Frequency-time distributions of earthquakes of different magnitude ranges in the four intervals of depth, i.e., 0-100 km, 100-300 km, 300-500 km and 500-700 km, are presented in Fig. 8. Note that in each particular graph the points from different colour lines at a given time represent the annual FM distribution of the Global seismic activity. Given frequency in logarithmic scale, the uniform distance between adjacent lines corresponds to the linear FM distribution, i.e., to the Gutenberg-Richter power law. The power-law exponent *b* can be derived from the value of this distance. Thus, by inspection of the graphs in Fig. 8 one can analyse temporal variations of the FM distribution, in particular, its slope and form. The following observations appear to be evident.

The behaviour of the FM graphs before 1995 is quite similar for all depths. The global seismic activity is rather stable and coherent at least above $mb = 5.0$ (while lower magnitudes are visibly incomplete). One may notice a short acceleration of activity in the mid-sixties apparently associated with the two greatest earthquakes in Alaska (1964) and Aleutian Islands (1965). During the following 30 years FM distributions are close to straight lines in magnitude range from 5.0 to 6.0. At shallow and intermediate depths these lines curve downwards at magnitudes above 6.0. The occurrences of strong earthquakes seem irregular. After 1995 a significant rearrangement of the FM distributions can be observed. It develops throughout the whole lithosphere but in a slightly different way for different depths.

At shallow depths it starts with the descent of moderate and strong earthquakes rates in 1996-1998, followed with acceleration, and continues in 2000-2004 with a steady rate and near uniform distance between neighbouring lines of the graph for all magnitudes $mb \geq 4.5$. Thus, the beginning of the $21^{st}$ century is characterised with the general straightening of the global FM distribution at shallow depths mainly due to the increase of the strong earthquake rate.

For depths 100-300 km and 300-500 km after 1995 the noticeable gradual diminution of seismic rate in the moderate magnitude range from 5.0 to 6.0 juxtaposes with the steady background rates of smaller and larger events. Such behaviour implies the convexity of the FM graph in its central part.

The FM graph at the ultra-deep layer after the 1994 shows a drastic redistribution of the seismic activity. Specifically, it comprises the evident growth of activity at $mb \geq 6.5$ ranges with the decrease of activity at moderate magnitudes from 6.4 to 5.0. These features correspond to the lowing down of the central part of the FM curve with its tail rising up.

Thus, from the viewpoint of *mb* frequency-magnitude distribution, during the first three decades or more of the period considered the global seismic activity follows rather steady process at all depths, then starting from about the mid-1990ies the evident transformation of the global seismic regime has occurred. It appears as gradual change of the global FM graphs from concave to convex. The most significant change occurred in the ultra-deep layer, below 500 km.

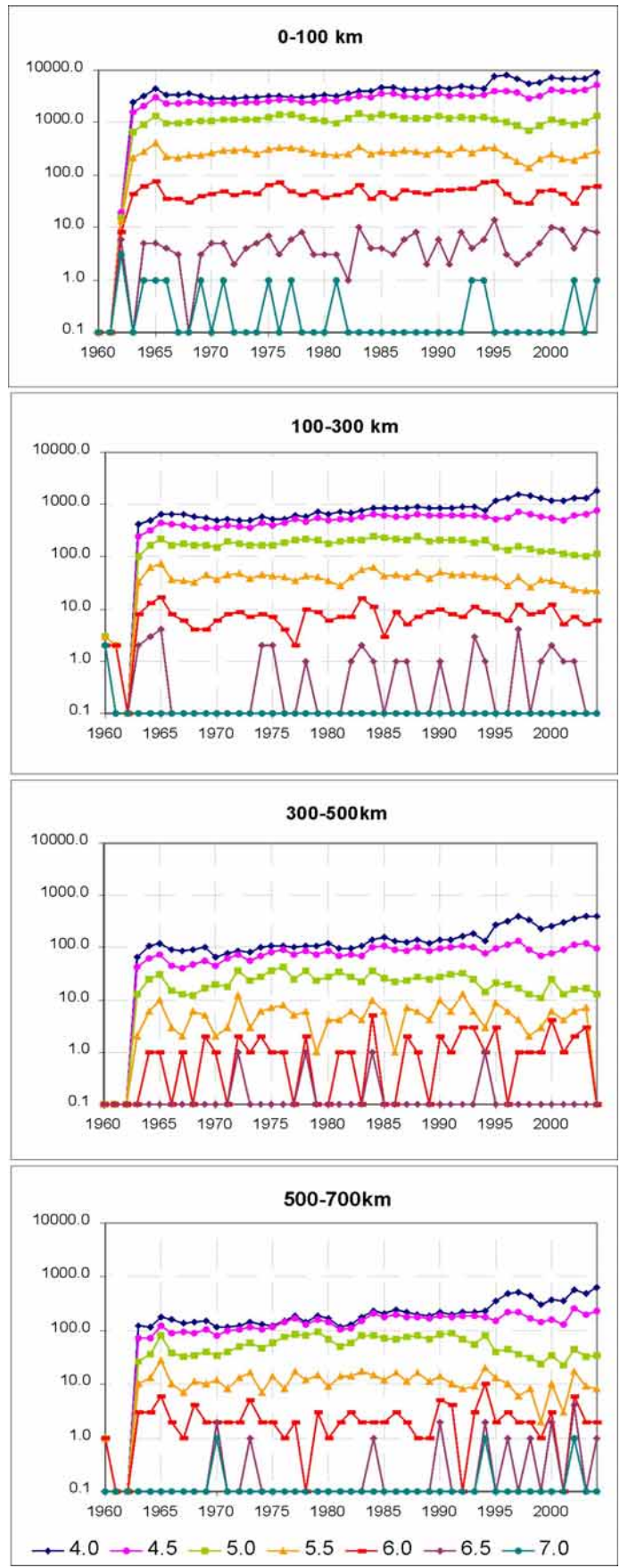

**Figure 8.** Annual number of earthquakes in the NEIC catalogue by time and *mb* magnitude, 1960-2004. Four diagrams correspond to the four depth intervals defined in Table 2.

Fig. 9 illustrates these observations with the empirical annual average FM distributions computed in the three 12-years intervals that subdivide 1969-2004, i.e., 1969-1980, 1981-1992, and 1993-2004. To facilitate cross comparison the graphs are given in pairs presented on the two plates. The first plate (Fig. 9a) shows the very similar behaviour at all depths in the first and the second intervals. The fit is rather impressive: the empirical FM curves have close slopes of the linear parts, indicate near equal levels of completeness, while differ by general levels of activity. Naturally, the shallow layer is the most active, followed by the intermediate and ultra deep populations of earthquake. The layer between 300 and 500 km is the quietest. The order is in accordance with the differential graphs presented in Fig. 3 above. For each specific depth the G-R graphs in 1969-1980 and 1981-1992 might be considered as almost identical. On the contrary, the third period reveals noticeably different form of the FM curves, especially, for intermediate, deep and ultra deep earthquakes (Fig. 9b). The lowing down of the central part of the FM curves around $mb$=5.0 is in common at all depths. Besides that in upper, ultra-deep and, perhaps, in deep layers the FM curve tails rise up.

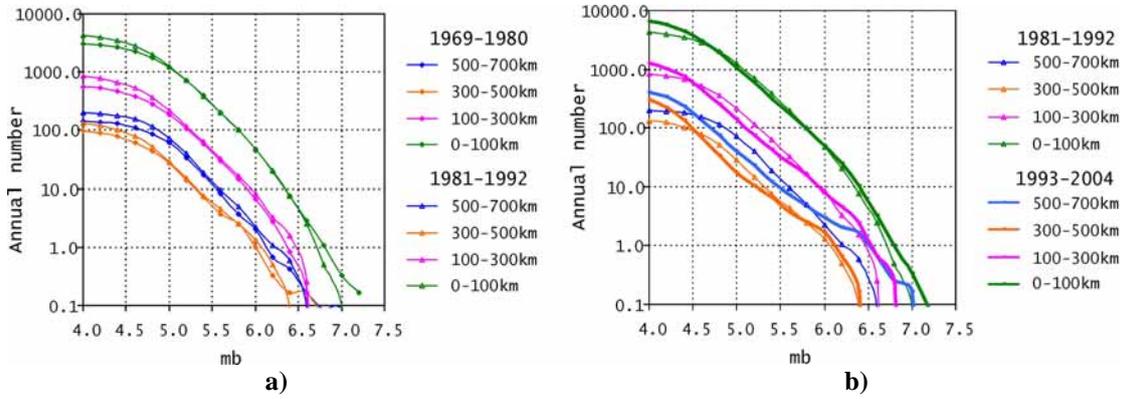

a)     b)

**Figure 9.** Annually averaged frequency-magnitude distributions for $mb$, calculated for pairs of consecutive time periods: a) 1969-1980 and 1981-1991; b) 1981-1992 and 1993-2004. Different coloured lines represent different depths.

To study the intermediate-term dynamics of the FM distribution curves we analyse temporal variations of their slope, coefficient $b$, evaluated on consecutive segments of the magnitude range. Specifically, the linear fit approximation is used:

$$b = (\log_{10}(N(M_1)) - \log_{10}(N(M_2))/\Delta M, \qquad (2)$$

where $N(M_i)$ is the number of events with magnitude $M_i$ and above, $\Delta M = M_2 - M_1 = 0.5$. Trailing intervals of 6 years shifted by 6 months are considered. The accepted 6-years duration is large enough to guarantee abundant statistics of earthquakes and is small enough to resolve the observable intermediate-term variations within the 36 years of the catalogue available. Besides that such span has given a good account of itself in the on-going real-time earthquake prediction experiment (Kossobokov et al. 1999; Romashkova & Kossobokov, 2005). The results for the specified depths are presented in Fig. 10. Note that the tick-marks on abscissa point at the end of 6-year intervals, so that the ordinates indicate $b$-value to-date. For example, the $b$-values at January 1, 2000 is computed from the earthquake populations in the period from January 1, 1994 through December 31, 1999.

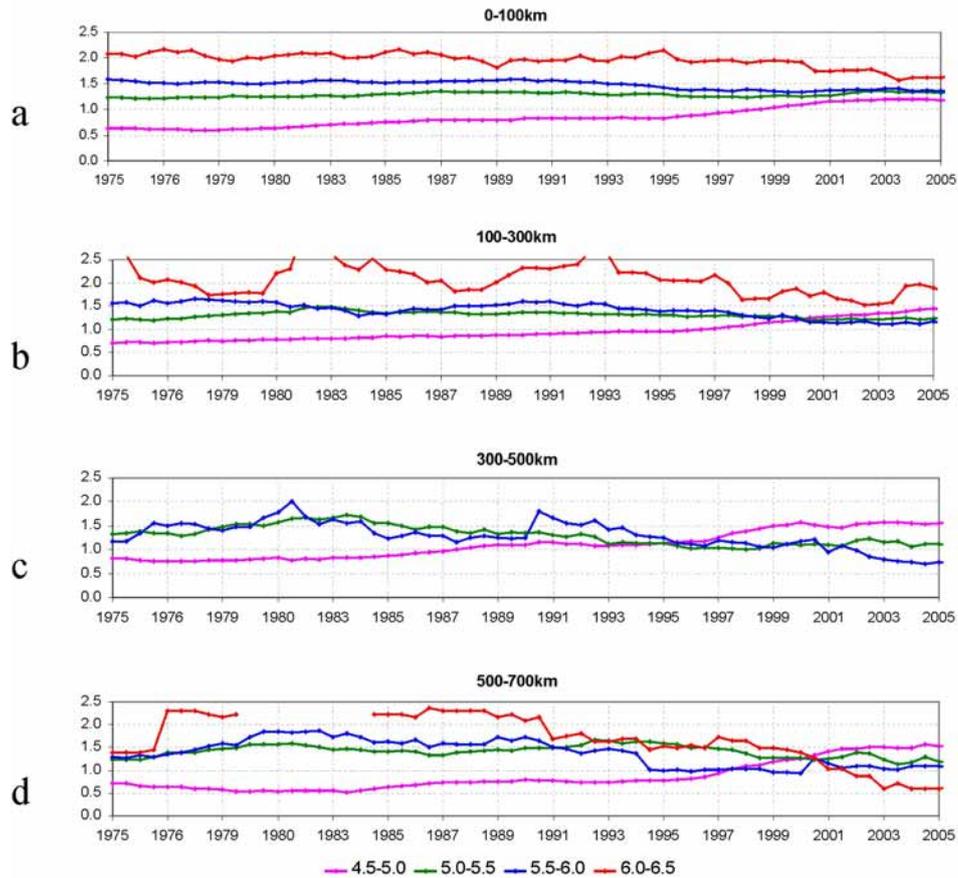

**Figure 10.** Time variations of b-value of the FM curves calculated for *mb* for several depth bands: a) 0-100km, b) 100-300km, c) 300-500km, d) 500-700km. Different coloured lines correspond to the different magnitude segments of FM curve. Time axis ticks mark the end of 6 years sliding interval. See text for detailed description.

The results summarised in Fig. 10 can be described as follows:

- **Depths 0-100 km**. In the first two decades of the analysis the FM curve is essentially concave with the *b*-value changing gradually from $b \approx 0.6$ at magnitudes *mb*=4.5-5.0 (violet line) up to *b* above 2.0 at *mb*=6.0-6.5 (red line). The values of *b* about 0.6 are apparently due to certain incompleteness of data at low magnitude ranges in the first decades. The extreme values of *b* above 2.0 characterise the small number of strong, $mb \geq 6.0$ events and fast saturation of the FM plots that involve an *mb*-type scale. For magnitudes *mb*=5.0-5.5 (green line) $b \approx 1.3$ and remains stable during the whole period. For *mb*=5.5-6.0 (blue line) the *b*-value declines from $b \approx 1.5$ starting at 1995 or earlier and reaches the level of $b \approx 1.3$ in 2002, since which time the FM curve fits a near straight line for *mb*=5.0-6.0. In general, after 2000 the whole FM curve tends to straighten up with its central part, $b \approx 1.35$.
- **Depths 100-300 km**. The FM curve starts from the concave form with some periods of linearity at segment of moderate magnitudes (green and blue lines) with a common *b*-value about 1.4. The tail of the FM curve at large *mb* values (red line) is rather unstable due to the small sample statistics and goes up and down frequently. Starting from 2000 the FM curve becomes convex for *mb*=4.5-6.0, with $b \approx 1.2$ for *mb*=5.0-5.5. The tails of the FM curve at extreme values of *mb* do not rise.

- **Depths 300-500 km**. Due to deficiency of data the FM curve is unstable at high magnitudes, while at moderate magnitudes (green and blue lines) it looks pretty linear up to 2001 although with the common slope varying between 1.5 and 1.0. The lower segment (violet line) has similar slope from the early 1990ies. After 2001 the FM curve is evidently convex with the $b$-value changing from $b \approx 1.5$ at low magnitudes to $b \approx 0.7$ at high magnitudes.
- **Depths 500-700 km**. Through the early 1990ies the FM curve appears essentially concave with the slope $b \approx 1.5$ at $mb=5.0-5.5$, then it gradually straightens up and becomes convex after 2001 with the slope varying from $b \approx 1.5$ at low to $b \approx 0.6$ at high values of $mb$.

Thus, one may conclude that starting from about late nineties the FM curves based on $mb$ attributed to intermediate, deep and ultra-deep earthquakes change their form from concave to convex within the magnitude range from 4.5 to 6.0, or even 6.5 in ultra-deep layer. For earthquakes from the upper layer general straightening of the FM curve is observed.

**Table 3**. Averaged over 10-year periods $b$-value for $mb$ magnitude on the interval [5.0; 6.0] of the FM curve, with standard deviation $\sigma$. Four consecutive depth bands are presented.

| Depth range/ | 0 – 100 km | | 100 – 300 km | | 300 – 500 km | | 500 – 700 km | |
|---|---|---|---|---|---|---|---|---|
| Period | \<b\> | $\sigma$ | \<b\> | $\sigma$ | \<b\> | $\sigma$ | \<b\> | $\sigma$ |
| 1975-1984 | 1.39 | 0.02 | 1.43 | 0.05 | 1.51 | 0.15 | 1.51 | 0.16 |
| 1985-1994 | 1.43 | 0.03 | 1.42 | 0.04 | 1.36 | 0.09 | 1.51 | 0.06 |
| 1995-2004 | 1.33 | 0.03 | 1.25 | 0.07 | 1.01 | 0.08 | 1.19 | 0.07 |

**Table 4**. Averaged over 10-year periods the concavity parameter *CON* for $mb$ magnitude on the interval [5.0; 6.0] of the FM curve, with standard deviation $\sigma$. Four consecutive depth bands are presented.

| Depth range/ | 0 – 100 km | | 100 – 300 km | | 300 – 500 km | | 500 – 700 km | |
|---|---|---|---|---|---|---|---|---|
| period | \<con\> | $\sigma$ | \<con\> | $\sigma$ | \<con\> | $\sigma$ | \<con\> | $\sigma$ |
| 1975-1984 | 0.07 | 0.01 | 0.05 | 0.04 | 0.01 | 0.04 | 0.04 | 0.03 |
| 1985-1994 | 0.05 | 0.01 | 0.04 | 0.02 | 0.01 | 0.06 | 0.01 | 0.06 |
| 1995-2004 | 0.02 | 0.01 | 0.00 | 0.00 | -0.02 | 0.05 | -0.07 | 0.04 |

Tables 3 and 4 summarise the time behaviour of the FM curve for moderate magnitude range $mb=5.0-6.0$. Specifically Table 3 gives the consecutive ten-year averages and standard deviations of the $b$-values determined by formula (2) for each of the four layers of depth. The overlapping six-year periods shifted by half-year are used for the $b$-value estimates. Tables 3 suggests that the average $b$-values in the first two decades are close to each other, while those in the third decade, i.e., in 1995-2004, differ significantly from the two previous (being separated by 2.1-3.5 standard deviations). It concerns most of all the deep and ultra-deep earthquake populations.

Table 4 provides the consecutive ten-year averages and standard deviations of the parameter *CON* determined as follows:

$$CON(M_1, M_2) = \log_{10} N((M_1+M_2)/2) - (\log_{10} N(M_1) + \log_{10} N(M_2))/2 \qquad (3)$$

in the same time windows as the *b*-value estimates. The parameter *CON* represents some measure of deviation of the FM curve from the straight line at the middle of the magnitude interval considered. The positive *CON*-values indicate concave form of the curve, while the negative – convex form. Table 4 suggests gradual change of the FM curve from concave to straight line in the upper and intermediate layers or to convex in the deep and ultra-deep layers.

Thus, a general conclusion can be drawn: *Starting from the mid- nineties the global mb-based FM distribution of earthquakes has experienced the qualitative transformation: In the upper-depth layer it appears as general straightening of the FM curve, while in the intermediate, deep and ultra-deep ones it changes the FM curve from concave to convex.* The first is a result of the gradual increase of the rate of the strong ($mb \geq 6.5$) shallow earthquakes, while the second starts with the decrease of the rate of moderate ($5.0 \leq mb < 6.0$) earthquakes in the early nineties, and follows by the rise of the rate of strong earthquakes.

**NEIC, $M_S$**  Fig. 1b shows the evolution of the annual rate of earthquakes from different ranges of $M_S$. Since only a few events deeper than 50 km hold an $M_S$ determination in NEIC the curves for shallow events are very close. One may grasp a similar pattern in the behaviour of the curves as the one observed with *mb* data at shallow depths in 1995-2004. In fact, the pattern of gradual increase of the strong earthquake rate on the background of stable or reduced rate of moderate earthquakes, resulting a near equal distance between different colour lines on the graph, is even more evident for $M_S$ data. As demonstrated above for *mb*-based plots, such behaviour corresponds to the straightening of the FM curve owing to the rise of the curve tail. The effect for $M_S$ is clearly observed in Fig. 11.

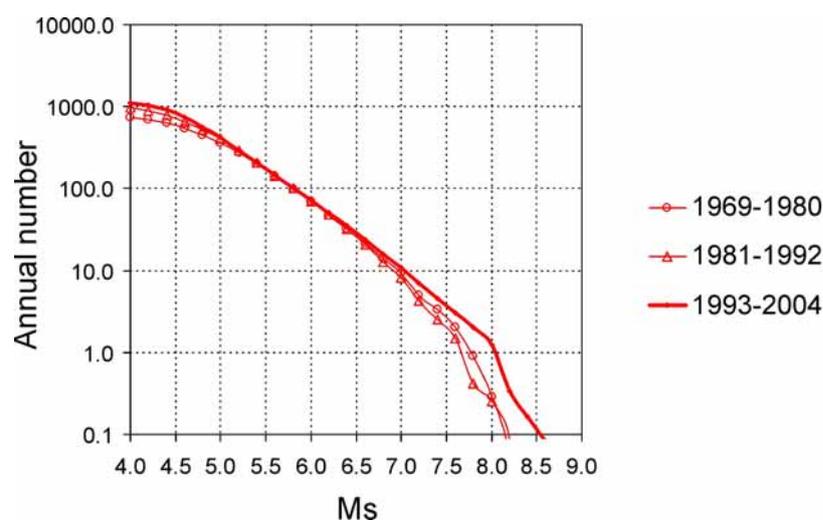

**Figure 11.** Annually averaged frequency-magnitude distributions for $M_S$, for three consecutive time periods.

Fig. 12 displays the intermediate-term dynamics of the $b$-value of the $M_S$ FM curve (computed in the same way as for $mb$) for several magnitude segments with $\Delta M$=1.0. The observed focusing of all $b$-value curves into a narrow beam after 2001 evidences the straight-line type of the FM distribution over the wide range of $M_S$ from 5.0 to 8.0. The common $b$-value is close to 0.8. Note that such pattern is rather atypical for $M_S$ data: in the period considered it covers less than 15% of the total. Usually the FM curve looks either concave, as in 1982-2000, or near linear with a steep falloff, as in 1979-1981. In the same way as in Table 3 for $mb$, Table 5 sums up the $b$-values observed in the three magnitude segments of the $M_S$ FM curve and consecutive 5-year periods. The FM slopes of the first and the second magnitude segments are rather stable, varying about $b$=0.74 for $M_S$=5.0-6.0 and about $b$=0.88 for $M_S$ =6.0-7.0. According to Figs 1b and 4, moderate $M_S$ data might be incomplete before 1980, which implies a relative underestimation of the $b$-values. The slope at $M_S$=7.0-8.0 has changed in the past 30 years from 1.66 to 0.87; the main falloff occurred in the last five (possibly, ten) years.

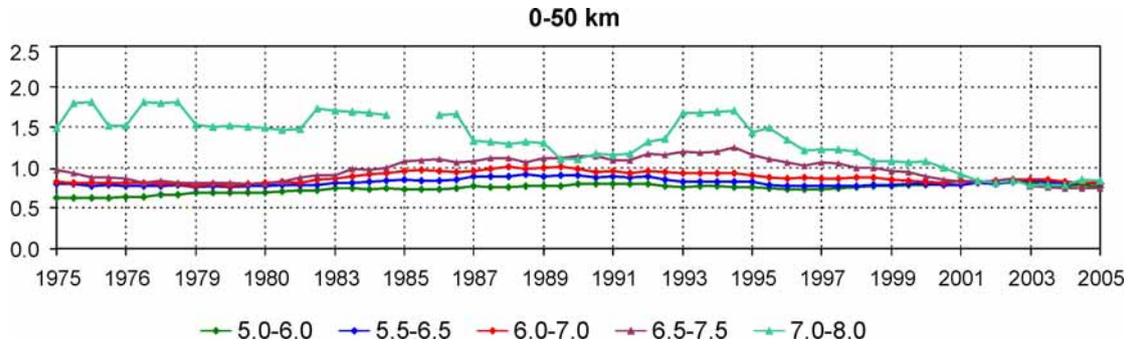

**Figure 12.** Time variations of b-value for $Ms$. Different coloured lines correspond to the different magnitude segments of FM curve. Time axis ticks mark the end of 6 years sliding interval.

**Table 5**. Averaged over 5-year periods $b$-values for $Ms$ magnitude on set of magnitude segments of the FM curve, with standard deviation $\sigma$.

| $Ms_{NEIC}$ range/ | 5.0 – 6.0 | | 6.0 – 7.0 | | 7.0 – 8.0 | |
|---|---|---|---|---|---|---|
| time period | <b> | $\sigma$ | <b> | $\sigma$ | <b> | $\sigma$ |
| 1975 – 1979 | 0.65 | 0.03 | 0.80 | 0.01 | 1.66 | 0.16 |
| 1980 – 1984 | 0.72 | 0.02 | 0.85 | 0.05 | 1.59 | 0.11 |
| 1985 – 1989 | 0.76 | 0.02 | 0.98 | 0.02 | 1.37 | 0.19 |
| 1990 – 1994 | 0.78 | 0.02 | 0.95 | 0.02 | 1.40 | 0.26 |
| 1995 – 1999 | 0.75 | 0.02 | 0.87 | 0.02 | 1.23 | 0.15 |
| 2000 – 2004 | 0.80 | 0.02 | 0.83 | 0.02 | 0.87 | 0.10 |

Thus, the analysis of the $M_S$ determinations contributes to the conclusion derived above from the $mb$ ones: *The last decade is marked with a general increase in the rate of the major $M_S \geq 7.0$ earthquakes world-wide on the background of stationary rate at moderate magnitude ranges; after 2001 the Gutenberg-Richter plot is linear with $b \approx 0.8$ over the entire range of reliable values of $M_S$ from 5.0 to 8.0.*

**CMT, $M_W$**   Similar to Fig. 9, Fig. 13 displays the FM curves computed from the CMT $M_W$ data in the two consecutive 12-year periods. A substantial uplift of the FM curve at high values of $M_W$ in the second period is evident for the upper and ultra-deep layers. Some sagging of the FM plot at moderate values of $M_W$ can be recognised for earthquakes from deep and, more clearly, from ultra-deep layers.

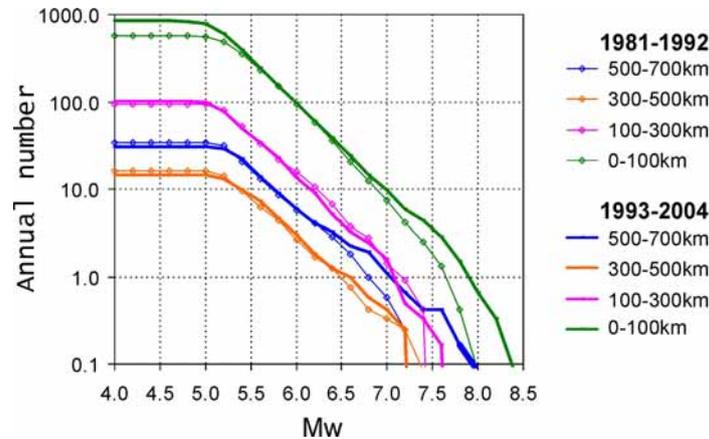

**Figure 13.** Annually averaged frequency-magnitude distributions for $Mw$ calculated for two of consecutive time periods: 1981-1992 (thin lines) and 1993-2004 (solid lines). Different colours represent different depth bands

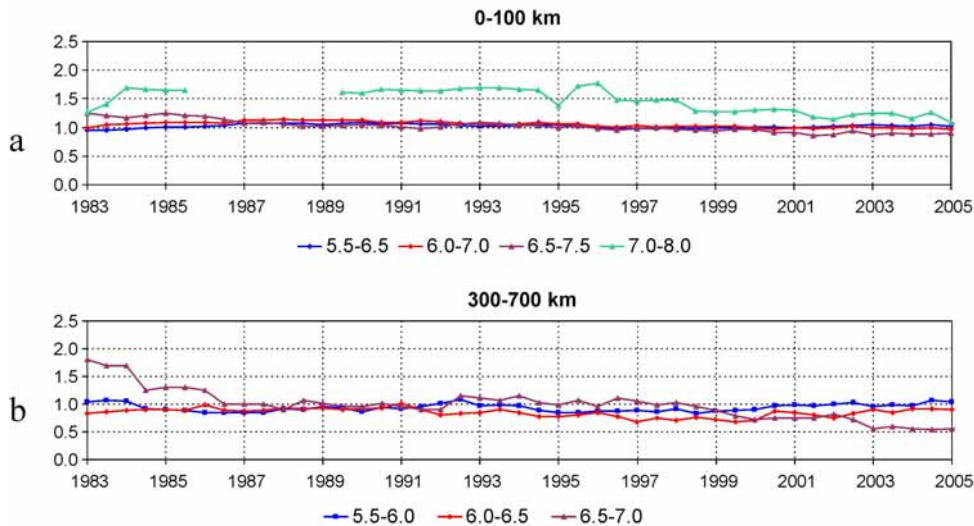

**Figure 14.** Time variations of b-value of $Mw$ FM curves calculated for different depth bands: (a) 0-100 km; (b) 300-700km. Different coloured lines correspond to the different magnitude segments of FM curves. Time axis ticks mark the end of 6 years sliding interval.

Fig. 14 shows the intermediate-term dynamics of the *b*-value of the $M_W$ FM plots determined in the two layers: shallow depth down to 100 km and deep from 300 to 700 km (due to a small number of earthquakes with reported $M_W$ in ultra-deep layer it is considered together with the deep one). As could be judged from the figure, the $M_W$ FM curve for shallow earthquakes fits a straight line at $M_W$=5.5-7.0 through the entire period of analysis. The *b*-value is close to 1.0, following the original moment magnitude scale definition (Kanamory, 1977) up to high values, where the $M_W$ FM curve has a falloff, which slope is not steady. The $M_W$ scale saturation for major and great shallow earthquakes might be attributed to the limited width of Earth crust that implies changing earthquake source geometry around $M_0=10^{27.2}$ dyne·cm (Okal & Romanowicz, 1994). As a result the $M_W$ FM curve is expected to band from *b*=1.0 to $b=1.5$ at $M_W$ about 7.4. In our analysis the $b$-value average of the $M_W$ FM curve for shallow earthquakes at $M_W=7.0$-$8.0$ over 1983-1997 equals 1.6, and it drops to about 1.2 in 1998-2004. This is an evidence of straightening of the FM curve from 1998 on.

In 1983-1985, the FM curve for deep earthquakes is visibly banded at $M_W=6.5$-$7.0$, while in 1986-2002 it is close to a straight line with a varying $b$-value between 0.7 and 1.0; then, starting from 2003, the curve becomes convex, uplifted at high values with the $b$-value approaching 0.5.

Thus, similar to $mb$ and $M_S$ scales, the analysis of the $M_W$ data shows that *recently the global $M_W$ frequency-magnitude distribution for shallow earthquakes has transformed towards a straight line and has become convex for deep and ultra-deep ones*.

## The two corollary evidences

Our analysis of global seismic activity shows its intermediate-term variability including significant deviations from quasi-stationary regime and qualitative conversions of the frequency-magnitude distributions, which could be viewed as characteristic anomalies. The observed anomalies did appear recently, after 1994 or later. They developed coherently throughout the lithosphere depths and had similar manifestations when different magnitude scales are considered. This suggests natural inferences on the character of the global processes in the lithosphere of the Earth.

(1) The observed temporal variations of the FM curves at different depths evidence *the global redistribution of seismic energy release in the lithosphere during the last years*. Specifically, besides the general increase of the total number of largest earthquakes world-wide, a considerable imbalance in the ratio between deep and shallow earthquake rates has occurred at that time. One can see (Fig. 15) that after 1994, and especially after 2002, the ratio of the largest, $mb \geq 6.5$, ultra-deep events total to that of the shallow ones (violet line) increases sizeably. The ratio reaches its maximum 21% in 2003 at which time it exceeds average level of 4.1% registered before 1994 by a factor of five. On the contrary, in the last years the ratio calculated for the moderate, $mb \geq 5.0$, earthquakes (green lines) decreases steadily from 6% to 3%. Similar behaviour of the ratios is obtained for depths from 300 to 700 km.

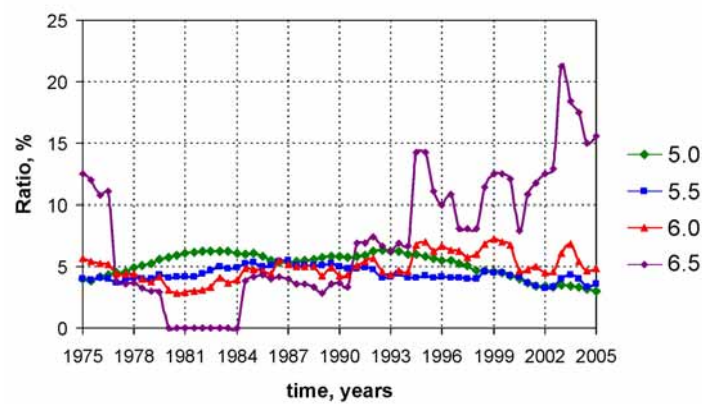

**Figure 15.** Time variation of ratio (in %) of the ultra-deep (500-700 km) to upper-deep (0-100 km) earthquake number in NEIC. Different coloured lines provide ratio for earthquakes above a certain magnitude threshold *mb*. (Notes: 1) The calculations are performed using the trailing 6-year windows. 2) The observed sexennial averages of ultra-deep events of magnitude 6.5, 6.0, 5.5 and 5.0 or larger are 3, 15, 70 and 365, correspondingly, while those of shallow depth earthquakes are 31, 292, 1603 and 7039.)

As can be seen from Fig. 15 there is another period, 1975-1976, featuring similar, although not so vivid, imbalance in the ratio. The character of this deviation is uncertain due to the short truncated span of the pattern: it may be either a short fluctuation in seismic distribution or a final, decaying part of the global rise of seismic activity in the middle of the 20th century. Note that these two periods differ a lot in the form of the FM distributions presented earlier in Fig. 10.

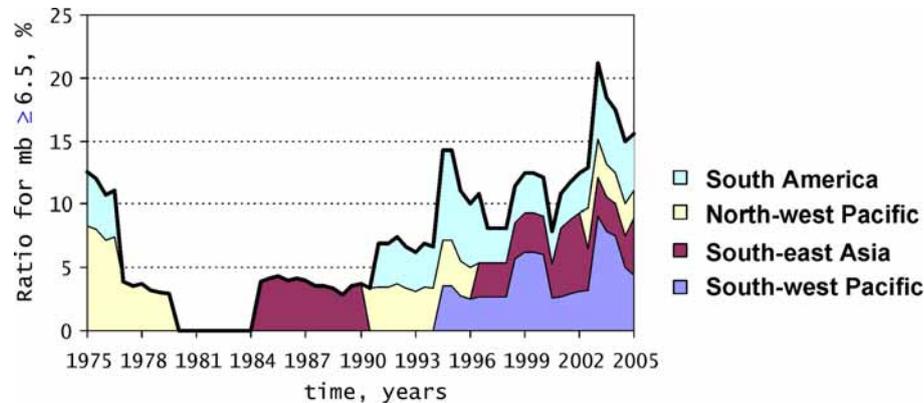

**Figure 16.** Distribution of ratio (in %) of the ultra-deep (500-700 km) to upper-deep (0-100 km) earthquake number, $mb \geq 6.5$, among regions. Different coloured bands provide contributions of different regions: SAm – South America, NWP – north-west Pacific, SEA – south-west Asia, SWP – south-west Pacific.

Fig. 16 investigates whether the observed pattern is of global nature by presenting the break-down of the ratio in proportion to the contributions from the four regions of deep earthquakes with hypocenter depths of 500 km or more. These regions are: (i) South America, Sam; (ii) North-West Pacific, NWP, including Kamchatka, Kuril islands, Japan and Bonin-Mariana trench; (iii) South-East Asia, SEA, including Indonesia and Philippine islands; (iv) South-West Pacific, SWP, including Tonga-Kermadec trench, and New Guinea. In fact an increase of the ultra-deep earthquake activity usually arise from one or two individual regions. It was so for a long time up to 1994. Then three regions, in different combinations, were showing up and after 2002 to the present the ultra-deep seismic activity spreads over all the four regions, each of which gives more or less equal contribution to the total (SAm – 2, NWP – 1, SEA – 2, SWP – 2 events in 6 years). Thus, it is possible to conclude that (i) *the redistribution of seismic energy release at depths occurred in the last decade is the global-scale one, and* (ii) *it combines the correlated worldwide rise of the ultra-deep activity manifested by great earthquakes with a relative reduction of the background seismic rate at shallow depths*.

(2) Another conclusion is pertinent to the cross-relations between different magnitude scales suggesting that these relations reveal a qualitative change in the last decade.

We have analysed the intermediate-term variations of the difference between $M_W$ and $mb$ (or $M_S$) scales for shallow (0-50 km) and deep (300-700 km) earthquakes separately. The average difference between $M_W$ and either $mb$ or $M_S$ has been calculated for moderate earthquakes ($mb \geq 5.0$, $M_S \geq 5.0$) from consequent magnitude bins $\Delta M_W=0.5$ within sliding 6-year interval. The temporal variations of the differences are shown in Fig. 17. From 1984 and up to 1995 $M_W - mb$ for shallow (a) and deep (b) earthquakes and magnitude range 5.5-7.0 (blue, red and violet lines) appears to indicate validity of a linear approximation $M_W = C_1 \cdot mb + C_2$ with $C_1 > 1$. Much larger values of the difference for earthquakes with $mb = 7.0$ or more (sea-green line) are due to the well-known saturation of the *mb*-scale. The variation of $M_W$ – *mb* and even its values are very similar in the two depth ranges. Fig. 17b evidences transformation of the graphs after 1995. On the background of steady rise of the 5.5-6.0 graph from its minimum in 1988 to the maximum in 2002, the transformation starts with the 6.0-6.5 graph moving down in 1995, then, in 2001, follows with the 6.5-7.0 graph moving also down and eventually crossing the 6.0-6.5 graph. Thus, the three graphs converge and $C_1$ approaches 1 from about 2002. On the contrary, shallow earthquakes preserve the linear relationship after 1995 (Fig. 17a), which varies slowly in the longer-term.

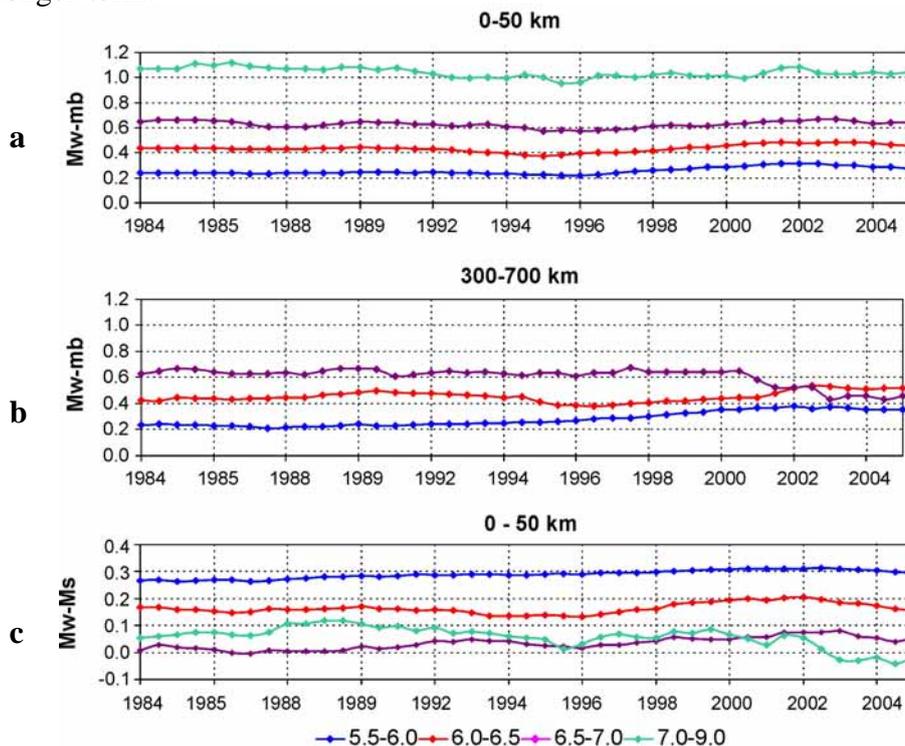

**Figure 17.** Temporal variations of the average difference between *Mw* and other magnitudes: a) *Mw* – *mb*, shallow events, b) *Mw* – *mb*, deep events, c) *Mw* – *Ms*, shallow events. Different coloured lines correspond to the different magnitude bins of *Mw*. Time axis ticks mark the end of 6 years sliding interval.

Similar to that the difference $M_W - M_S$ available for shallow earthquakes only shows up the validity of a linear approximation $M_W = C_1 \cdot M_S + C_2$ with $C_1 < 1$ in the $M_S$ magnitude range 5.5-7.0 (Fig. 17c). For the strongest, 7.0-9.0, earthquakes (sea-green line) the graph is not much in line with the others until 2003 when it dives below 0 resulting the linear approximation valid trough the entire magnitude range considered, i.e., 5.5-9.0.

Thus, it is possible to conclude that *the qualitative transformation of the cross-relations between different magnitude scales took place globally during the last decade or even few years*.

## Discussion and conclusions

The variations of the global seismicity were subject of constant interest of seismologists at all times (Benioff 1951; Mogi 1979; Romanowicz 1993; Bufe & Perkins 2005). Usually researches exploit different concepts of seismic cycle and succeed observing some global scale patterns in seismic energy release over decades.

The general global approach to the Earth seismic dynamics naturally arises from the conception of the Earth lithosphere as a complex non-linear dynamic system consisting of hierarchy of interacting different-scale blocks (Keilis-Borok, 1990, 2002; Keilis-Borok & Soloviev, 2003). The lithosphere as a whole represents an ultimate member of this hierarchy. Seismic activity is a manifestation of its internal processes, most of which remain "invisible" to geoscientists. Accordingly, each mega-earthquake represents an extreme critical event of the manifestation. The results of present study demonstrate that the lithosphere does behaves in a way typical for such systems and reveals classical symptoms of instability in advance the 2004 Sumatra-Andaman mega-earthquake, at least at the intermediate-term scale of years. These symptoms, systematically explored, depicted by different patterns of seismic activity and tested for earthquake prediction purposes by many authors, are based on (i) *variation of seismic activity*, (ii) *its redistribution in space*, (iii) *changes in frequency-magnitude distribution*, (iv) *cross-relations between different magnitude scales*, and (v) *range of spatial and temporal correlation.*

The present work demonstrates primarily *rise* of the global rate of strong earthquakes in the last decade, which has induced upward bending of the frequency-magnitude graph at the highest values for all magnitude scales and all depths. Moreover, this rise is accompanied by a relative drop in the rate of moderate earthquakes at depths below 100 km in particular. Such a drop could be recognised as precursory *quiescence*. Therefore, the global seismic activity in the last decade illustrates that both premonitory patterns may coexist in time and space when definitions of rise and quiescence refer to different magnitude ranges (or scales). The coexistence of rise and drop of activity were not the same at different depths that results the coherent world-wide redistribution of seismic energy release.

The observed changes of the global frequency-magnitude graphs are in line with predictive modelling and laboratory experiments aimed at better understanding of fractures and fracturing (Scholz, 1968; Smith, 1981; Knopoff et al., 1982; Meredith et al., 1990; Molchan & Dmitrieva, 1990; Barenblatt, 1993; Narkunskaya & Shnirman, 1994). In particular, similar premonitory phenomena have been recognized by Rotwain et al. (1997) who (i) noted that in laboratory the break up of a sample is preceded by transition from active formation of new micro-cracks to their coalescence or expansion into macro- or even major-cracks and (ii) reported such a transformation of seismic activity in Southern California. In terms of the frequency-magnitude relation the observed changes mean that at the first stage *b*-value increases at smaller magnitudes then drops at larger magnitudes. This results in the central part of the FM graph going down and shaping the entire graph concave. The same behaviour has been demonstrated in the present study for global deep and ultra-deep seismicity a decade prior to the 2004 Sumatra-Andaman mega-earthquake.

Cross-relations between different magnitude scales are closely related to the earthquake source geometry and physical properties of its environment. It is of common knowledge that *mb* is measured at about 1 s, $M_S$ - at 20 s (Aki & Richards, 1980), and $M_W$ – presumably (but not necessarily) through all periods. For *mb* and $M_S$ this results into the magnitude scale saturation at high values when earthquake source linear size exceeds wave-length used for the magnitude determination. The difference of definitions allows using the routine determinations of magnitudes for estimation of the source spectral content (Keilis-Borok, 1960; Prosorov & Hudson, 1974). The analysis of the $M_S$ – *mb* statistic (Kaveraina et al., 1996) allowed a conclusion that predominance of high frequencies in the source spectrum is connected with relatively small earthquake source size and shot duration of rupture process, which may indicate high level of tectonic stress and relatively low temperature of the source environment. The saturation of $M_W$ for shallow earthquakes is likely related to earthquake source geometry (Okal & Romanowicz, 1994). All these considerations permit hypothesising that temporal changes of cross-relations between different magnitude scales apparently reflect physical changes in the lithosphere, and may indicate either approach to the critical point or stable state of the system.

Many dynamic systems show up coherent behavior at the approach of extreme event (Thom, 1975; Prigogine & Stengers, 1984), which could be expressed in *increase of correlation range*. The concept of "long-range correlation" was introduced in seismology by A. Prozorov (1975) when he examined remote, so-called "long-range aftershocks" of major earthquakes as indicators of the future ones' location. Other examples of long-range correlation include premonitory wide-spread occurrence of near-simultaneous earthquakes (Shebalin et al., 2000) and rise of seismic activity in sufficiently large number of fault zones (Zaliapin et al, 2002). A convincing demonstration of precursory collective behavior of seismic activity over a large area before strong earthquakes is given by real-time applications of earthquake prediction algorithms (Kossobokov et al. 1999; Rotwain & Novikova, 1999; Keilis-Borok & Soloviev, 2003). The algorithms make use of a whole series of seismic patterns that arise coherently within an area of linear size 5-10 times the size of target earthquake. As has been shown above, the lithosphere as a whole reveals patterns of collective behaviour a decade before the 2004 Sumatra-Andaman mega-earthquake. These global scale patterns are similar to those detected earlier on a regional scale before many great, major, and strong earthquakes.

The results of our study suggest (i) *the presence of global scale intermediate-term tectonic processes in the lithosphere*; (ii) *the occurrence of the global scale seismicity patterns implying criticality in the last decade*. The December 26, 2004 Sumatra-Andaman mega-earthquake may represent either the culmination or one of the successive stages of such a process in the last decade. A better understanding of the problem could be achieved in comparative analysis of geophysical environment preceding and following other greatest earthquakes. Note that in our study we did not investigate short-term characteristics of the global seismic activity. Neither did we consider other than splitting by depth possibilities of spatial break-down. Both deserve consideration in further research. General studies aimed at finding typical features for damped stages of critical phenomena distinct to local slowdowns of on-going process may facilitate such understanding. In case of extreme seismic events, geophysical data other than seismic catalogues could be useful and may become critical for understanding the physics of the global processes occurring in the Earth interior and, eventually, may help predicting seismic catastrophes like mega-earthquakes.

## Acknowledgements

The research has been partly supported by the Program "Change in Environment and Climate: Natural Catastrophes" of the Presidium of the Russian Academy of Sciences, and EC Sixth Framework Programme (Priority FP6-2003-NEST-PATH "Tackling Complexity in Science", Project E2-C2 "Extreme Events: Causes and Consequences").

We are grateful to G. Molchan and V. Keilis-Borok for valuable discussion and advice in preparing the manuscript.

# References

Aki, K. & Richards, P.G., 1980. Quantitative seismology: theory and methods, *W.H.Freeman and Company,* San Francisco.

Barenblatt, G.I., 1993. Micromechanics of fracture, *Elsvier*, TUTAM, 25-51.

Barenblatt, G.I., Keilis-Borok, V.I., Monin, A.S., 1983. Filtration model of earthquake sequence. *Dokl. Akad. Nauk SSSR* 269(4): 831-834 (in Russian).

Benioff, V.H., 1951. Global strain accumulation and release as revealed by great earthquakes, *Bull. Geol. Soc. Am.* 62, 331-338.

Bowman, D.D., Ouillon, G., Sammis, C.G., Sornette, A., Sornette, D., 1998. An observational test of the critical earthquake concept, *J. Geophys. Res.* 103: 24359-72.

Dobrovolsky, I.P., Zubkov, S.I., Miachkin, V.I., 1979. Estimation of the size of earthquake preparation zone, *Pure Appl. Geoph.*,117, No.5, 1025-1044.

Ekstrom, G., A. M. Dziewonski, N. N. Maternovskaya, M. Nettles, 2005. Global seismicity of 2003: centroid-moment tensor solutions for 1087 earthquakes, *Phys. Earth Planet. Inter.*, 148, 327–351.

GHDB (Global Hypocenter Data Base), 1989. *Global Hypocenters Data Base CD-ROM NEIC/USGS*, Denver, CO, 1989 and its updates through December 2004.

Gutenberg, B. & Richter, C.F., 1954. Seismicity of the Earth and associated phenomena, *Princeton Universiy Press,* Princeton, N. J.

Kagan Y.Y. & Jackson D.D., 1991. Long-term earthquake clustering, *Geophys. J. Int.,* 104, 117-133.

Kanamory, H., 1977. Energy release in great earthquakes, *J. Geophys. Res.,* 82, 2981-2987.

Kaverina, A.N., Prozorov, A.G., Lander, A.V, 1996. Global creepex distribution and its relation to earthquake-source geometry and tectonic origin, *Geophys. J. Int.*, 125, 1, 249-265.

Keilis-Borok, V.I., 1960. The difference between spectrums of surface waves of earthquakes and underground explosions, *Trans. of the Inst. Phys. Earth,* 15 (in rissian).

Keilis-Borok, V.I., 1990. The lithosphere of the Earth as a nonlinear system with implications for earthquake prediction, *Rev. Geophys*. 28, 1: 19-34.

Keilis-Borok. V.I., 1994. Symptoms of instability in a system of earthquake faults. *Physica D,* 77, 193-199.

Keilis-Borok. V.I., 2002. Earthquake prediction: state-of-the-art and emerging possibilities. *Annu. Rev. Earth Planet. Sci.,* 30, 1-33.

Keilis-Borok, V.I. & Malinovsaya, L.N., 1964. One regularity in the occurrence of strong earthquakes, *J. Geopgys. Res.,* 69, 3019-3024.

Keilis-Borok, V.I., & A.A. Soloviev, (Editors), 2003. Nonlinear Dynamics of the Lithosphere and Earthquake Prediction, *Springer, Heidelberg*, 348 p.

Knopoff, L., Kagan, Y., Knopoff, R., 1982. B-value for foreshocks and aftershocks in real and simulated earthquake sequences, *Bull. Seismol. Soc. Am.,* 72, 1663-1676.

Kossobokov, V.G., V.I. Keilis-Borok, L.L. Romashkova, and J.H. Healy, 1999. Testing earthquake prediction algorithms: Statistically significant real-time prediction of the largest earthquakes in the Circum-Pacific, 1992-1997, *Phys. Earth Planet.Inter.*, 111, 3-4, 187-196.

Meredith P.G., Main I.G., Jones, C., 1990. Temporal variations in seismicity during quasistatic and dynamic rock failure, *Tectonophysics,* 175, 249-268.

Mogi, K., 1968. Migration of seismic activity, *Bull. Earth, Res. Inst. Univ. Tokyo,* 46(1), 53-74.
Mogi, K., 1979. Global variation of seismic activity, *Tectonophysics,* 57, T43-T49.
Mogi, K., 1985. Earthquake prediction, *Academic Press,* Tokyo.
Molchan, G.M. & Dmitrieva, O.E., 1990. Dynamics of the magnitude-frequency relation for foreshocks, *Phys. Earth Planet.Inter.*, 61, 99-112.
Narkunskaya G.S. & Shnirman, M.G., 1994. An algorithm of earthquake prediction, *Computational Seismology and Geodynamics,* Vol.1, AGU, Washington, D.D., 20-24.
Okal, E.A. & Romanowicz, B.A., 1994. On the variation of b-values with earthquake size. *Phys. Earth Planet.Inter.*, 87, 55-76.
Prigogine, I, & I. Stengers, 1984. Order out of Chaos. Bantam Books, NY.
Press, F. & Allen C., 1995. Patterns of seismic release in the southern California region, *J.Geophys. Res.,* 100, No. B4, 6421-6430.
Prozorov, A.G., 1975. Changes of seismic activity connected to large earthquakes, *Interpretation of Data in Seismology and Neotectonics, Comput. Seismol.,* 8, 71-82 (in Russian).
Prozorov, A.G. & Hudson, J.A., 1974. A study of the magnitude difference Ms–mb for earthquakes, *Geophys. J. R. Astr. Soc.,* 39, 551-554.
Prozorov, A.G., Schreider, S.Y., 1990. Real time test of the long-range aftershocks algorithm as a tool of mid-term earthquake prediction in South California, *Pure Appl. Geophys.,* 133, N.1, 1-19.
Romanowicz, B., 1993. Spatiotemporal patterns in the energy release of great earthquakes, *Science,* 260, 1923-1926.
Romashkova, L. & Kossobokov, V., 2005. Earthquake prediction based on stable clusters of alarms. *Abstracts of the Contributions of the EGU 2-st General Assembly*, Vienna, Austria, EGU05-A-02065.
Rotwain, I., Keilis-Borok, V., Botvina, L., 1997. Premonitory transformation of steel fracturing and seismicity, *Phys. Earth Planet.Inter.*, 101, 61-71.
Rotwain, I.M. & Novikova, O.V, 1999. Performance of the earthquake prediction algorithm CN in 22 regions of the world, *Phys. Earth Planet. Int.,* 111, 207-213.
Sadovsky M.A, editor, 1986. Long-term earthquke prediction: methodological recommendations, *Inst. Phys. Earth, Moscow* (in Russian).
Scholz, C.H., 1968. The frequency-magnitude relation of microfracturing in rocks and its relation to earthquakes, *Bull. Seismol. Soc. Am.,* 58, N 1, 399-415.
Shebalin, P.N., 1992. Automatic duplicate identification in set of earthquake catalogues merged together, *U.S. Geol. Surv. Open-File Report 92-401*, Appendix II.
Shebalin P., Zaliapin I., Keilis-Borok V., 2000. Premonitory raise of the earthquake's correlation range: Lesser Antiles, *PEPI* 122, 241-249.
Smith, W.D., 1981. The b-value as an earthquake precursor, *Nature,* 289, N 5794, 136-139.
Sornette, D. & Sammis, 1995. Complex critical exponents from renormalization group theory of earthquakes: Implication for earthquakes predictions, *J. Phys. I France*, 5: 607-619.
Sykes, L.R., and Jaume, S.C., 1990. Seismic activity on neighboring faults as a long-term precursor to large earthquakes in the San Francesco Bay area, *Nature,* 348, 595-599.
Thom, R., 1975. Structural stability and morphogenesis, Benjamin Reading, Massachusetts.

Turcotte, D.L. Newman, W.I. & A. Gabrielov, 2000. A statistical physics approach to earthquakes. In *Geocomplexity and the Physics if Earthquakes, American Geophysical Union*, Washington D.C.

Vil'kovich, E.V. and Shnirman M.G., 1983. Epicenter migration waves: Examples and models. In V.I. Leilis-Borok and A.L. Levshin, editors, *Mathematical Models of the Earth's Structure and the Earthquake Prediction, Comput. Seism.* 14, 27-36, Allerton Pres, New York.

Wallace, R.E., 1987. Grouping and migration of surface faulting and variation in slip rates on faults in Greate Basin province, *Bull. Seism. Soc. Am.,* 77, 868-876.

Zaliapin, I., Keilis-Borok, V., and Axen G., 2002. Premonitory spreading of seismicity over the faults' network in southern California: Precursor Accord, *J.Geophys. Res.,* 107, No. B10, 2221.